\documentstyle[amssymb,preprint,prd,aps]{revtex}

\global\arraycolsep=2pt
\makeatletter
\def\fmslash{\@ifnextchar[{\fmsl@sh}{\fmsl@sh[0mu]}}
\def\fmsl@sh[#1]#2{  \mathchoice
    {\@fmsl@sh\displaystyle{#1}{#2}}    {\@fmsl@sh\textstyle{#1}{#2}}    
{\@fmsl@sh\scriptstyle{#1}{#2}}    {\@fmsl@sh\scriptscriptstyle{#1}{#2}}}
\def\@fmsl@sh#1#2#3{\m@th\ooalign{$\hfil#1\mkern#2/\hfil$\crcr$#1#3$}}
\makeatother

\begin{document}
\draft
\title{Criterion for local distinguishability of arbitrary bipartite orthogonal
states }
\author{Ping-Xing Chen$^{1,2}${\footnotesize \thanks{%
E-mail: pxchen@nudt.edu.cn}} and Cheng-Zu Li$^1$}
\address{1. Department of Applied Physics, National University of\\
Defense Technology,\\
Changsha, 410073, \\
P. R. China. \\
2. Laboratory of Quantum Communication and Quantum Computation, \\
University of Science and Technology of\\
China, Hefei, 230026, P. R. China}
\date{\today}
\maketitle

\begin{abstract}
In this paper we present a necessary and sufficient condition of
distinguishability of bipartite quantum states. It is shown that the
operators to reliably distinguish states need only rounds of projective
measurements and classical comunication. We also present a necessary
condition of distinguishability of bipartite quantum states which is simple
and general. With this condition one can get many cases of
indistinguishability. The conclusions may be useful in understanding the
essence of nonlocality and calculating the distillable entanglement and the
bound of distillable entanglement.
\end{abstract}

\pacs{PACS number(s): 03.67.-a, 03.65.ud }

\thispagestyle{empty}

\newpage \pagenumbering{arabic} 

One of the interesting features of non-locality in quantum mechanics is that
a set of orthogonal quantum states cannot be distinguished if only a single
copy of these states is provided and only local operations and classical
communication (LOCC) are allowed, in general. The procedure of
distinguishing quantum states locally is: Alice and Bob hold a part of a
quantum system, which occupies one of $m$ possible orthogonal states $\left|
\Psi _1\right\rangle ,\left| \Psi _2\right\rangle ,...,\left| \Psi
_i\right\rangle ,...,\left| \Psi _m\right\rangle $. Alice and Bob know the
precise form of these states, but don't know which of these possible states
they actually hold. To distinguish these possible states they will perform
some operations locally: Alice (or Bob) first measures her part. Then she
tell the Bob her measurement result, according to which Bob measure his
part. With the measurement results they can exclude some possibilities of
the system\cite{1}. Briefly speaking, the procedure of distinguishing
quantum states locally is to exclude all or some possibilities by
measurement on the system. Many authors have considered some schemes for
distinguishing locally between a set of quantum states \cite{1,2,22,3,4,44,5}%
, both inseparable and separable. Bennett et al showed that some orthogonal
product states cannot be distinguished by LOCC\cite{2}. Walgate et al showed
that any two states can be distinguished by LOCC\cite{1}. For two-qubit
systems (or $2\otimes 2$ systems), any three of the four Bell states:

\begin{eqnarray}
\left| A_1\right\rangle &=&\frac 1{\sqrt{2}}(\left| 00\right\rangle +\left|
11\right\rangle )  \label{1} \\
\left| A_2\right\rangle &=&\frac 1{\sqrt{2}}(\left| 00\right\rangle -\left|
11\right\rangle )  \nonumber \\
\left| A_3\right\rangle &=&\frac 1{\sqrt{2}}(\left| 01\right\rangle +\left|
10\right\rangle )  \nonumber \\
\left| A_4\right\rangle &=&\frac 1{\sqrt{2}}(\left| 01\right\rangle -\left|
10\right\rangle )  \nonumber
\end{eqnarray}
cannot be distinguished by LOCC if only a single copy is provided\cite{3}.
The distinguishability of quantum states has some close connections\cite{61}
with distillable entanglement\cite{62} and the information transformation%
\cite{63}. On one hand, using the upper bound of distillable entanglement,
relative entropy entanglement\cite{52} and logarithmic negativity\cite{53},
the authors in Ref \cite{3} proved that some states are indistinguishable.
On the other hand, using the rules on distinguishability one can discuss the
distillable entanglement\cite{61}. So the further analysis for
distinguishability is meaningful. In this paper, we will give a necessary
and sufficient condition of distinguishability of bipartite quantum states.
It is shown that the operators to reliably distinguish states need only
rounds of projective measurements and classical comunication. We also
present a necessary condition of distinguishability of bipartite quantum
states which is simple and general. With this condition one can get many
cases of indistinguishability.

Consider $m$ possible orthogonal states shared between Alice and Bob. Any
protocol to distinguish the $m$ possible orthogonal states can be conceived
as successive rounds of measurement and communication by Alice and Bob. Let
us suppose Alice is the first person to perform a measurement (Alice goes
first \cite{22}), and the first round measurement by Alice can be
represented by operators $\left\{ A_{1_j}\right\} $, where $%
A_{1_j}^{+}A_{1_j}$ is known as a POVM element realized by Alice \cite{h,p},
and $\sum_jA_{1_j}^{+}A_{1_j}=I.$ If the outcome $1_j$ occurs, then the
given $\left| \Psi \right\rangle $ becomes $A_{1_j}\left| \Psi \right\rangle
,$ up to normalization. After communicating the result of Alice's
measurement to Bob, he carries out a measurement and obtain outcome $1_k$.
The given possible state $\left| \Psi \right\rangle $ becomes $%
A_{1_j}\otimes B_{1_k}(1_j)\left| \Psi \right\rangle $, where $B_{1_k}(1_j)$
is an arbitrary measurement operator of Bob which depend on the outcome $1_j$
of Alice's measurement. After N rounds of measurements and communication,
there are many possible outcomes which corresponding to many measurement
operators acting on the Alice and Bob's Hilbert space. Each of these
operators is a product of the N sequential and relative operators $%
A_{1_j}\otimes B_{1_k}(1_j)A_{2_j}(1_j,1_k)\otimes
B_{2_k}(1_j,1_k,2_j)...A_{N_j}(1_j,1_k,...,(N-1)_j)\otimes
B_{N_k}(1_j,1_k,...,(N-1)_j,N_k)$ carried out by Alice and Bob. We denote
these operators as $\left\{ A_i\otimes B_i\right\} ,$ where, $A_i\otimes $ $%
B_i$ denotes one of these operators, which represent the effects of the N
measurements and communication. If the outcome $i$ occurs, the given $\left|
\Psi \right\rangle $ becomes:

\begin{equation}
A_i\otimes B_i\left| \Psi \right\rangle
\end{equation}
The probability $p_i$ Alice and Bob gain outcome $i$ is

\begin{equation}
p_i=\left\langle \Psi \right| A_i^{+}\otimes B_i^{+}A_i\otimes B_i\left|
\Psi \right\rangle ,
\end{equation}
and

\begin{equation}
\sum_iA_i^{+}\otimes B_i^{+}A_i\otimes B_i=I
\end{equation}
Suppose we define:

\begin{equation}
E_i=A_i^{+}\otimes B_i^{+}A_i\otimes B_i,
\end{equation}
then $E_i$ is a positive operator and that $\sum_iE_i=I.$ $E_i$ is similar
to the POVM element $A_i^{+}A_i.$ We can regard $E_i$ as a generalized POVM
(GPOVM) element, which has same property as known POVM element. In fact, $%
A_i $ can be written in the form $A_i=U_{A2}f_{Ai}U_{A1}\ $\cite{p}$,$ where 
$f_{Ai}$ is a diagonal positive operator, $U_{A2},U_{A1}$ are two unitary
operators, and similarly for $B_i.$ If each of $N$ Alice's operators denoted
by $A_i$ and each of $N$ Bob's operators denoted by $B_i$ are projectors,
the final operators $A_i\otimes B_i$ are also projectors, i.e., $A_i\otimes
B_iA_j\otimes B_j=\delta _{ij}A_i\otimes B_i$, and $\left\{ E_i\right\} $ is
a set of projective measurement.

The discuss above means that: whatever Alice and Bob choose to do, including
they decide to involve an ancillary system; they perform local unitary
operators and measurement; they use one-way or two-way communication, and do
many rounds of measurements and communication, their final actions will be
described by a set positive operators $\left\{ E_i\right\} .$ The
probability of a given possible state $\left| \Psi \right\rangle $ yielding
a certain outcome $i$ is $\left\langle \Psi \right| A_i^{+}\otimes
B_i^{+}A_i\otimes B_i\left| \Psi \right\rangle .$

Since a GPOVM element $E_i$ has similar property as a POVM element, $E_i$
can be represented in the form

\begin{eqnarray}
E_i &=&(a_1^i\left| \varphi _1^i\right\rangle _A\left\langle \varphi
_1^i\right| +\cdots +a_{m_a}^i\left| \varphi _{m_a}^i\right\rangle
_A\left\langle \varphi _{m_a}^i\right| +\cdots )\otimes  \label{e} \\
&&(b_1^i\left| \eta _1^i\right\rangle _B\left\langle \eta _1^i\right|
+\cdots +b_{m_b}^i\left| \eta _{m_b}^i\right\rangle _B\left\langle \eta
_{m_b}^i\right| +\cdots )  \nonumber \\
0 &\leqslant &a_{m_a}^i\leqslant 1,0\leqslant b_{m_b}^i\leqslant
1;1\leqslant m_a\leqslant N_a,1\leqslant m_b\leqslant N_b
\end{eqnarray}
where $\left\{ \left| \varphi _{m_a}^i\right\rangle \right\} ,\left\{ \left|
\eta _{m_b}^i\right\rangle \right\} $ is a set of bases of Alice's and
Bob's, respectively; $N_a,N_b$ is the dimensions of Alice's and Bob's
Hilbert space, respectively.

Theorem 1. If a set of $m$ orthogonal states $\left\{ \left| \Psi
_i\right\rangle \right\} $ is reliably locally distinguishable, if and only
if there is a set of locally distinguishable product vectors, $\left\{
\left| \varphi _i^{j_i}\right\rangle _A\left| \eta _i^{j_i}\right\rangle
_B\right\} ,$ such that each state $\left| \Psi _i\right\rangle $ can be
written as a superposition of some different product vectors in the set of 
these product vectors:

\begin{equation}
\left| \Psi _i\right\rangle =c_i^1\left| \varphi _i^1\right\rangle _A\left|
\eta _i^1\right\rangle _B+\cdots +c_i^{n_i}\left| \varphi
_i^{n_i}\right\rangle _A\left| \eta _i^{n_i}\right\rangle _B
\end{equation}
where $i=1,...,m$; $j_i=1,...,n_i.$ $n_i$ is the number of product bases in
the state $\left| \Psi _i\right\rangle $

Proof: The proof of sufficiency is obviously. Now we prove the necessarity.
If a set of states is reliably locally distinguishable, there must be a set
of GPOVM element $\left\{ E_i\right\} $ such that if every outcome $i$
occurs Alice and Bob know with certainty that they were given the state $%
\left| \Psi _i\right\rangle $. In a simple way we say that $E_i$ can and
only can ``indicate'' $\left| \Psi _i\right\rangle .$

Since each $E_i$ only indicate a state $\left| \Psi _i\right\rangle ,$ the
rank of $E_i$ should be less than $N_aN_b$. Otherwise, $E_i$ will indicates
all states $\left\{ \left| \Psi _i\right\rangle \right\} $. Without loss of
generality, we suppose $a_1^i,...,a_{m_a}^i,b_1^i,...,b_{m_b}^i$ in Eq. (\ref
{e}) are nonzero, the other coefficient are zero, then the state $\left|
\Psi _i\right\rangle $ should have all or part of the component 
\begin{equation}
\left| \varphi _1^i\right\rangle _A\left| \eta _1^i\right\rangle _B,\cdots
,\left| \varphi _1^i\right\rangle _A\left| \eta _{m_b}^i\right\rangle
_B,\cdots ,\left| \varphi _{m_a}^i\right\rangle _A\left| \eta
_1^i\right\rangle _B,\cdots ,\left| \varphi _{m_a}^i\right\rangle _A\left|
\eta _{m_b}^i\right\rangle _B.
\end{equation}
The state$\left| \Psi _j\right\rangle $ $(i\neq j)$ should have not these
component. Otherwise, $E_i$ will indicates more than one state. The effect
of the operator $E_i$ is project out the component

\begin{equation}
\left| \varphi _1^i\right\rangle _A\left| \eta _1^i\right\rangle _B,\cdots
,\left| \varphi _1^i\right\rangle _A\left| \eta _{m_b}^i\right\rangle
_B,\cdots ,\left| \varphi _{m_a}^i\right\rangle _A\left| \eta
_1^i\right\rangle _B,\cdots ,\left| \varphi _{m_a}^i\right\rangle _A\left|
\eta _{m_b}^i\right\rangle _B
\end{equation}
which only belongs to $\left| \Psi _i\right\rangle .$ Because of the
completeness of $\left\{ E_i\right\} $ ( which assures that each component
in all possible states can be indicated by a GPOVM element) and the
necessarity of reliably distinguishing the possible states (which askes a
GPOVM element only indicate the component only belonging to a possible
states), the $m$ possible states can be divided into many component each of
which can be indicatd by a GPOVM element. It is to say that these component
are locally distinguishable. If a operator $E_i$ only indicate a state, then 
$E_i$ can be replaced by a set of operators

\begin{equation}
E_{i1}=\left| \varphi _1^i\right\rangle _A\left\langle \varphi _1^i\right|
\otimes \left| \eta _1^i\right\rangle _B\left\langle \eta _1^i\right| ;...; 
\nonumber
\end{equation}
\begin{equation}
E_{im_b}=\left| \varphi _1^i\right\rangle _A\left\langle \varphi _1^i\right|
\otimes \left| \eta _{m_b}^i\right\rangle _B\left\langle \eta
_{m_b}^i\right| ;...;
\end{equation}
\[
E_{im_am_b}=\left| \varphi _{m_a}^i\right\rangle _A\left\langle \varphi
_{m_a}^i\right| \otimes \left| \eta _{m_b}^i\right\rangle _B\left\langle
\eta _{m_b}^i\right| , 
\]
each of which also only indicates the same state as $E_i$ does. The effect
of each operator $E_{ij}(j=1,...,m_am_b)$ is to project out a component
which only belong to a single state, for example, operator project out the
component $\left| \varphi _1^i\right\rangle _A$ $\left| \varphi
_1^i\right\rangle _B$. Thus all product vectors which only belong to a state
can be indicated by a est of operators $\left\{ E_{ij}\right\} ,$ so all the
product vectors are locally distinguishable and then are orthogonal. Because
of the completeness of $\left\{ E_i\right\} $ and the necessarity of
reliably distinguishing the possible states, each state $\left| \Psi
_i\right\rangle $ must be the superposition of some locally distinguishable
orthogonal product bases. This end the proof.

From theorem 1 it follows that if a set of operators $\left\{ E_i\right\} $
can distinguish a set of states, there is a set of orthogonal projective
operators $\left\{ E_{ij}\right\} $ which can achieve the distinguishing the
possible states. These operators can always be carried out by rounds of
projective measurements and classical communication. It is to say that if a
set of $m$ states are distinguishable, the operator to distinguish the
states can be always described as: First, Alice and Bob choose a person to
go first to do measurement with her or his a set of orthogonal bases; After
measurement, their Hilbert space collapse into a subspace which is
orthogonal to the subpace discarded. According to the outcome, they know
that the $m$ possible states collapses into $n\prime $ ($n\prime \leqslant
n) $ distinguishable new states, and choose a person to do measurement with
her or his other set of orthogonal bases once more, and so on. After many
rounds of projective measurements and classical communication, they may get
a final state which only belongs to one of the possible states, and the
Hilbert space collapses into a lower subspace. The final state may be
entangled state or separable state. If the final state is entangled one can
choose end of the operators, also can choose continue to do measurement such
that it collapses into one of it's orthogonal product states. In each round
of measurement and communication, Alice and Bob must choose an appropriate
person to do measurement. The different round of operator many need
different person to do first, in general. For example, six states in a $%
4\otimes 4$ system,

\begin{equation}
\left| \Psi _1\right\rangle =\left| 0\right\rangle _A\left| 0\right\rangle
_B;\left| \Psi _2\right\rangle =\left| 1\right\rangle _A(\left|
0\right\rangle +\left| 1\right\rangle )_B;\left| \Psi _3\right\rangle
=\left| 0\right\rangle _A\left| 1\right\rangle _B+\left| 1\right\rangle
_A(\left| 0\right\rangle -\left| 1\right\rangle )_B
\end{equation}
\begin{equation}
\left| \Psi _4\right\rangle =\left| 2\right\rangle _A\left| 2\right\rangle
_B;\left| \Psi _5\right\rangle =(\left| 2\right\rangle +\left|
3\right\rangle )_A\left| 3\right\rangle _B;\left| \Psi _6\right\rangle
=\left| 3\right\rangle _A\left| 2\right\rangle _B+(\left| 2\right\rangle
-\left| 3\right\rangle )_A\left| 3\right\rangle _B
\end{equation}
one can first choose to do measurement with Alice's bases

\[
E_1=\left| 0\right\rangle _A\left\langle 0\right| +\left| 1\right\rangle
_A\left\langle 1\right| ;\quad E_1=\left| 2\right\rangle _A\left\langle
2\right| +\left| 3\right\rangle _A\left\langle 3\right| 
\]
if the outcome is $E_1$ they must choose Alice to do first to do the second
measurement; if the outcome is $E_2$ they must choose Bob to go first to do
the second measurement. The distinguishability of states in $2\otimes n$
systems is a special example in which after Alice do measurement, the
possible states collapse into some orthogonal product bases, so the possible
states can be written as the form in the Theorem 1 of Walgate's paper \cite
{22}.

Before giving theorem 2 in this paper, we define a concept of {\it Schmidt
number}. If a pure state $\left| \Psi \right\rangle $ have following Schmidt
decomposition:

\begin{equation}
\left| \Psi \right\rangle =\sum_{i=1}^l\sqrt{P_i}\left| \nu _i\right\rangle
_A\left| \eta _i\right\rangle _B,\qquad \qquad \qquad \sum_{i=1}^lP_i=1
\label{2}
\end{equation}
where $\left| \nu _i\right\rangle _A^{\prime }s$ and $\left| \eta
_i\right\rangle _B^{\prime }s$ are orthogonal bases of Alice and Bob,
respectively, we say $\left| \Psi \right\rangle $ has {\it Schmidt number} $%
l.$ Keep this in mind we will start from the following theorem.

Theorem 2: Alice and Bob share a quantum system, which possesses one of pure
orthogonal states $\left| \Psi _1\right\rangle ,\left| \Psi _2\right\rangle
,...,\left| \Psi _i\right\rangle ,...,\left| \Psi _m\right\rangle .$ If the
dimensions of Hilbert space of Alice's part and Bob's part are $N_a$ and $%
N_b,$ respectively, one cannot distinguish deterministically a set of
orthogonal states for which the sum of {\it Schmidt numbers} is more than $%
N_aN_b$ when only a single copy is provided.

Theorem 2 can be expressed briefly as: one cannot distinguish a set of
orthogonal states the sum of Schmidt number of which is more than the
dimensions of whole Hilbert space of the quantum system.

From theorem 2 one can get the following interesting cases:

Case1: For $n\otimes n$ systems one cannot distinguish deterministically $%
n+1 $ states, each of which has Schmidt number $n.$ For example, one can at
most distinguish two entangled states in $2\otimes 2$ systems.

Case 2: For $n\otimes n$ systems, if one can distinguish $n^2$ orthogonal
states, these states must be orthogonal bases .

Proof of theorem 2: If the $m$ possible orthogonal states are reliably
locally distinguishable there are a set of orthogonal product basis each of
which only emerges in one state of the $m$ possible states, and therefore
only ``indicates'' one state. A pure state with Schmidt numbers $l$ includes
at least $l$ orthogonal product bases, so if the sum of Schmidt numbers of
the $m$ possible states is more than $N_aN_b$, these possible states surely
include product bases more than $N_aN_b$. However, the number of orthogonal
product bases in a $N_a\otimes N_b$ system is at most $N_aN_b.$ So these $m$
possible states with Schmidt number more than $N_aN_b$ cannot be written as
the form in theorem 1, which means they are not distinguishable. This
completes the proof theorem 2.

According to the theorem 2 we can discuss completely the case for $2\otimes
2 $ systems in which there are at most four orthogonal states, as be shown
in Ref \cite{22}. Four orthogonal states can be distinguished
deterministically if and only if they are four product states. This is
obviously from the theorem above or case 2; three orthogonal states can be
distinguished deterministically if and only if almost one state is entangled
state and the other states are product states. This is because any two
product orthogonal states can be written as: $\left| \nu _1\right\rangle
\left| \eta _1\right\rangle $ and $\left| \nu _2\right\rangle \left| \eta
_2\right\rangle ,$ where at least one of the inner product $\langle \nu
_1\left| \nu _2\right\rangle $ and $\langle \eta _1\left| \eta
_2\right\rangle $ is zero. Without loss of generality, we let $\langle \nu
_1\left| \nu _2\right\rangle =0$, then $\left| \nu _1\right\rangle ,\left|
\nu _2\right\rangle $ consist of a complete set of orthogonal bases in two
dimensions Hilbert space. Any entangled state in $2\otimes 2$ systems can be
expressed as: $a\left| \nu _1\right\rangle \left| \eta _1^{^{\prime
}}\right\rangle +b\left| \nu _2\right\rangle \left| \eta _2^{^{\prime
}}\right\rangle .$ As these three states are orthogonal, the following
relation should hold: $\langle \eta _1\left| \eta _1^{^{\prime
}}\right\rangle =\langle \eta _2\left| \eta _2^{^{\prime }}\right\rangle =0.$
This means the three states are distinguishable deterministically. If more
than two states of the three orthogonal states are entangled, the sum of
their Schmidt numbers would be more than four, which does not satisfy the
necessary condition for distinguishability in the theorem above. So if and
only if not less than two states of three orthogonal states are product
states, the three states are distinguishable.

In summary, we present a necessary and sufficient condition of
distinguishability of bipartite quantum states. It is shown that the
operators to reliably distinguish states need only rounds of projective
measurements and classical communication.. We also present a necessary
condition of distinguishability of bipartite quantum states which is simple
and general. With this condition one can get many cases of
indistinguishability. These results come directly from the limits on local
operations, not from the upper bound of distillable entanglement\cite{3}, So
we believe that they may be useful in calculating the distillable
entanglement or the bound of distillable entanglement. The further works may
be the applications of these results.

\acknowledgments    We would like to thank professor M.B. Plenio for his
helpful suggestions and professor Guangcan Guo for his help to this work.

\end{document}